%% file: main.tex
\documentclass[a4paper,11pt]{article}
\usepackage{pos}
\usepackage[range-phrase=-, range-units=single]{siunitx}

\title{Probing neutrino emission at GeV energies from compact binary mergers detected during O1-O4a with the IceCube Neutrino Observatory}
\ShortTitle{GeV neutrinos from binary mergers with IceCube}

\author{The IceCube Collaboration \\{\normalsize \normalfont(a complete list of authors can be found at the end of the proceedings)}\\}

\emailAdd{karlijn.kruiswijk@uclouvain.be}
\emailAdd{mathieu.lamoureux@uclouvain.be}
\emailAdd{matthias.vereecken@uclouvain.be}
\emailAdd{gwenhael.dewasseige@uclouvain.be}

\abstract{Compact binary mergers, detected in gravitational waves since 2015, are candidate sources for astrophysical neutrinos in the GeV regime from proton-proton and proton-neutron collisions. This contribution presents the results of the search for such a signal using mergers detected during the fourth observing run of the LIGO, Virgo, and KAGRA interferometers. We use the dense infill array at the center of the IceCube detector, IceCube-DeepCore, to select neutrino candidates in the 0.5-–5 GeV energy range. The search for a statistically significant excess associated with an astrophysical signal is performed in a $\pm 500$\,s window around the gravitational wave detection time. We do not observe any statistically significant excess in the neutrino data, and set upper limits on the neutrino emission from these objects. Additionally, we search for subpopulations of neutrino-emitting sources, including merger events detected in previous observing runs; no significant signal has been identified yet.

\vspace{4mm}
{\bfseries Corresponding authors:}
Karlijn Kruiswijk, Mathieu Lamoureux$^{*}$, Matthias Vereecken, Gwenhaël de Wasseige\\
{\itshape Centre for Cosmology, Particle Physics and Phenomenology - CP3, Universit{\'e} Catholique de Louvain, Louvain-la-Neuve, Belgium}\\[4mm]
$^*$ Presenter}

\ConferenceLogo{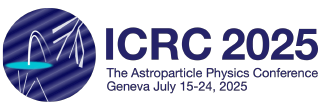}
\FullConference{39th International Cosmic Ray Conference (ICRC2025)\\ 15--24 July 2025\\ Geneva, Switzerland}

\begin{document}

\maketitle

\paragraph{Introduction}

Compact binary mergers, especially those involving at least one neutron star, are candidate neutrino emitters, especially in the GeV energy range~\cite{Murase:2013hh,Murase:2022}. Searches performed so far in this regime have been inconclusive~\cite{IceCube:2023atb,KM3NeT:2023cdr,Super-Kamiokande:2021dav}, reporting upper limits on the flux normalisation.

We report on the search for GeV neutrinos using the “Extremely-LOW-ENergy” (ELOWEN) sample, an event selection focusing on clusters of hits identified in DeepCore~\cite{IceCube:2011ucd}, the inner and denser sub-volume component of the IceCube Neutrino Observatory~\cite{IceCube:2016zyt}. The ELOWEN selection has a background rate of about $20-25$\,mHz and is mostly sensitive to neutrinos in the $0.5-5$\,GeV energy range.

More details on the motivation, context, methods, and results of the presented analysis can be found in the related publication~\cite{ELOWEN-GW}.

\paragraph{Analysis}

We focus on compact binary mergers detected in gravitational waves by LIGO-Virgo-KAGRA~\cite{LIGOScientific:2018mvr,LIGOScientific:2020ibl,LIGOScientific:2021usb,KAGRA:2021vkt} between the first observing run (O1, September 2015) and the first part of the fourth observing run (O4a, January 2024). For each gravitational wave event, we search for an excess of ELOWEN events in a $\pm 500$\,s time window centered on the gravitational wave detection time $t_{\rm GW}$. For mergers involving at least one neutron star, we also investigate a shorter time window $[t_{\rm GW}, t_{\rm GW} + 3\,\text{s}]$, as motivated notably by the observed time delay between the detection of the binary neutron star (BNS) merger GW170817 and the onset of the associated gamma-ray burst, GRB 170817A~\cite{LIGOScientific:2017ync}.

The analysis was performed for 178 sources, including 167 binary black hole (BBH) mergers, 9 neutron star -- black hole candidates (including the marginal candidate GW200105\_162426), and 2 BNS candidates. For each source, we compute the pre-trial p-value, and the best one is found to be $0.25\%$ for S230707ai. However, this corresponds to a $40\%$ post-trial p-value, therefore not significant.

\begin{figure}[hbtp]
    \centering
    \includegraphics[width=0.75\linewidth]{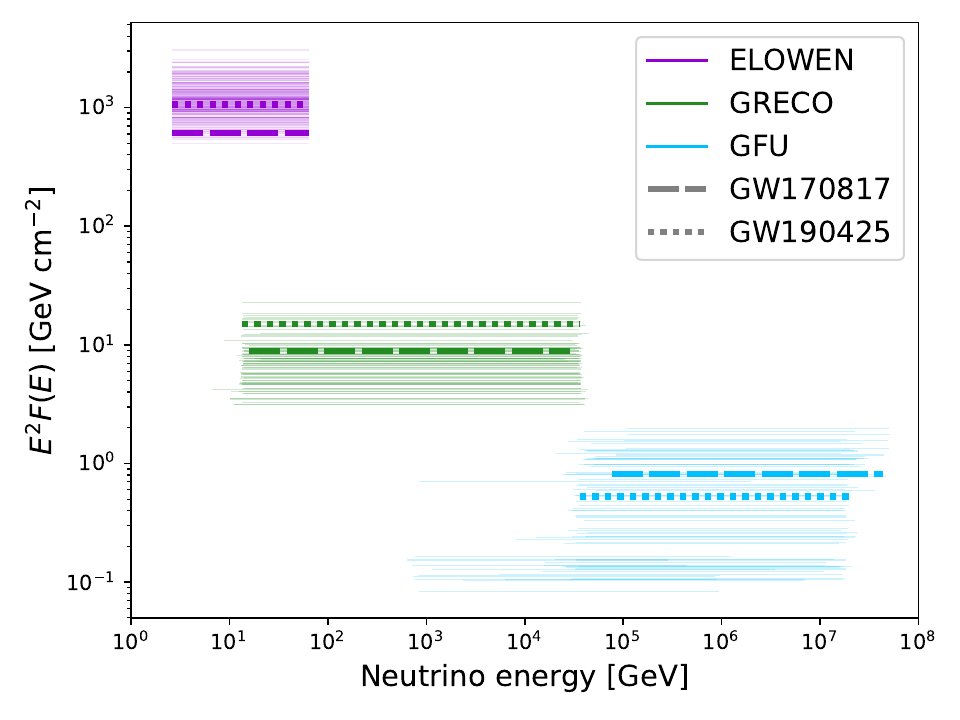}
    \caption{Summary of 90\% upper limits on the all-flavour time-integrated energy-squared flux $E^2 F_{\rm all-flavour}^{\nu+\bar\nu}(E)$ assuming a power-law spectrum $F_{\rm all-flavour}^{\nu+\bar\nu}(E) = \phi (E/\si{\GeV})^{-2}$. The different colours correspond to different IceCube samples: ELOWEN (this work), GRECO \cite{IceCube:2023atb}, and GFU \cite{IceCube:2022mma}. The spans in the horizontal axis correspond to the central 90\% sensitive energy range for each GW event and sample. The thicker dashed and dotted lines are the limits corresponding to the two confirmed BNS mergers, GW170817 and GW190425.}
    \label{fig:limits}
\end{figure}

In the absence of any excess, we compute upper limits on the all-flavour neutrino flux at Earth at 90\% credible level, assuming an $E^{-\gamma}$ spectrum, with $\gamma = 2, 2.5, 3$. \autoref{fig:limits} summarises the limits for $\gamma = 2$, compared to the results from the GRECO Astronomy~\cite{IceCube:2023atb} and the Gamma-ray Follow-up (GFU,~\cite{IceCube:2022mma}) datasets, which focus on GeV--TeV neutrinos and TeV--PeV track-like events, respectively.

Additionally, we perform a binomial test to search for evidence of subpopulations of neutrino emitters among the analysed mergers. We obtain a post-trial p-value of 81\%, fully consistent with the background hypothesis.

\paragraph{Conclusion}

Despite not being able to claim any significant detection, the presented analysis shows several promising prospects. First, as illustrated in \autoref{fig:limits}, the different IceCube event selections probe complementary energy ranges, therefore maximising the chances to eventually identify the underlying neutrino production mechanisms. The plot also highlights how joint analyses, combining different datasets, could strengthen constraints on the neutrino energy distribution. This approach could be extended in the future to include multiple neutrino telescopes, further improving source characterization.

Second, the limits will be significantly enhanced by new developments in the ELOWEN selection, which will improve the noise rejection and incorporate a limited direction reconstruction~\cite{IceCube:2023kjn}. Other approaches are currently under investigation to reduce the energy threshold even further.

Finally, the upcoming deployment of the IceCube Upgrade~\cite{Ishihara:2019aao}, notably including multi-PMT optical modules, will significantly improve the direction reconstruction capability and the effective area in the \si{\GeV} range, pushing further the sensitivity of future searches.

\bibliographystyle{JHEP}
\bibliography{references}

\clearpage

\input{authorlist_IceCube.tex}

\end{document}

%% file: authorlist_IceCube.tex
\section*{Full Author List: IceCube Collaboration}

\scriptsize
\noindent
R. Abbasi$^{16}$,
M. Ackermann$^{63}$,
J. Adams$^{17}$,
S. K. Agarwalla$^{39,\: {\rm a}}$,
J. A. Aguilar$^{10}$,
M. Ahlers$^{21}$,
J.M. Alameddine$^{22}$,
S. Ali$^{35}$,
N. M. Amin$^{43}$,
K. Andeen$^{41}$
% ...
\\
\\
% ...
$^{10}$ Universit{\'e} Libre de Bruxelles, Science Faculty CP230, B-1050 Brussels, Belgium \\
% ...
$^{16}$ Department of Physics, Loyola University Chicago, Chicago, IL 60660, USA \\
$^{17}$ Dept. of Physics and Astronomy, University of Canterbury, Private Bag 4800, Christchurch, New Zealand \\
% ...
$^{21}$ Niels Bohr Institute, University of Copenhagen, DK-2100 Copenhagen, Denmark \\
$^{22}$ Dept. of Physics, TU Dortmund University, D-44221 Dortmund, Germany \\
% ...
$^{35}$ Dept. of Physics and Astronomy, University of Kansas, Lawrence, KS 66045, USA \\
% ...
$^{39}$ Dept. of Physics and Wisconsin IceCube Particle Astrophysics Center, University of Wisconsin{\textemdash}Madison, Madison, WI 53706, USA \\
% ...
$^{41}$ Department of Physics, Marquette University, Milwaukee, WI 53201, USA \\
% ...
$^{43}$ Bartol Research Institute and Dept. of Physics and Astronomy, University of Delaware, Newark, DE 19716, USA \\
% ...
$^{63}$ Deutsches Elektronen-Synchrotron DESY, Platanenallee 6, D-15738 Zeuthen, Germany \\
$^{\rm a}$ also at Institute of Physics, Sachivalaya Marg, Sainik School Post, Bhubaneswar 751005, India \\
% ...

\subsection*{Acknowledgments}

\noindent
The authors gratefully acknowledge the support from the following agencies and institutions:
USA {\textendash} U.S. National Science Foundation-Office of Polar Programs,
U.S. National Science Foundation-Physics Division,
U.S. National Science Foundation-EPSCoR,
U.S. National Science Foundation-Office of Advanced Cyberinfrastructure,
Wisconsin Alumni Research Foundation,
Center for High Throughput Computing (CHTC) at the University of Wisconsin{\textendash}Madison,
Open Science Grid (OSG),
Partnership to Advance Throughput Computing (PATh),
Advanced Cyberinfrastructure Coordination Ecosystem: Services {\&} Support (ACCESS),
Frontera and Ranch computing project at the Texas Advanced Computing Center,
U.S. Department of Energy-National Energy Research Scientific Computing Center,
Particle astrophysics research computing center at the University of Maryland,
Institute for Cyber-Enabled Research at Michigan State University,
Astroparticle physics computational facility at Marquette University,
NVIDIA Corporation,
and Google Cloud Platform;
Belgium {\textendash} Funds for Scientific Research (FRS-FNRS and FWO),
FWO Odysseus and Big Science programmes,
and Belgian Federal Science Policy Office (Belspo);
Germany {\textendash} Bundesministerium f{\"u}r Forschung, Technologie und Raumfahrt (BMFTR),
Deutsche Forschungsgemeinschaft (DFG),
Helmholtz Alliance for Astroparticle Physics (HAP),
Initiative and Networking Fund of the Helmholtz Association,
Deutsches Elektronen Synchrotron (DESY),
and High Performance Computing cluster of the RWTH Aachen;
Sweden {\textendash} Swedish Research Council,
Swedish Polar Research Secretariat,
Swedish National Infrastructure for Computing (SNIC),
and Knut and Alice Wallenberg Foundation;
European Union {\textendash} EGI Advanced Computing for research;
Australia {\textendash} Australian Research Council;
Canada {\textendash} Natural Sciences and Engineering Research Council of Canada,
Calcul Qu{\'e}bec, Compute Ontario, Canada Foundation for Innovation, WestGrid, and Digital Research Alliance of Canada;
Denmark {\textendash} Villum Fonden, Carlsberg Foundation, and European Commission;
New Zealand {\textendash} Marsden Fund;
Japan {\textendash} Japan Society for Promotion of Science (JSPS)
and Institute for Global Prominent Research (IGPR) of Chiba University;
Korea {\textendash} National Research Foundation of Korea (NRF);
Switzerland {\textendash} Swiss National Science Foundation (SNSF).